\documentclass[a4paper,twocolumn,english,showpacs,nofootinbib,nobibnotes,aps,pra,floatfix,superscriptaddress]{revtex4-2}
\usepackage[utf8]{inputenc}
\usepackage{amsmath}
\usepackage{amsthm}
\usepackage{amssymb}
\usepackage{graphicx}
\usepackage[caption=false]{subfig}
\usepackage{xparse}
\usepackage{xcolor}
\usepackage{siunitx}
\usepackage{hyperref}
\usepackage{booktabs}
\usepackage{tabularx}

\usepackage{multirow}

\usepackage{times}
\usepackage{appendix}

\makeatletter

\usepackage{braket}
\usepackage{dsfont}

\makeatother

\usepackage{babel}

\global\long\def\ketbra#1{\ket{#1}\!\bra{#1}}

\newcommand{\kommentar}[1]{}

\NewDocumentCommand\opti{smmm>{\SplitList{;}}m} {
\begingroup%
\setlength{\belowdisplayskip}{-0.6\baselineskip}%
\IfBooleanTF{#1}{%
    \begin{alignat*}{2}
        & \underset{#3}{\text{#2}} & & #4 \\
        & \text{subject to~~}
        \ProcessList{#5}{ \insertopticonst }
        & &
    \end{alignat*}%
    }{%
    \begin{alignat}{2}
        & \underset{#3}{\text{#2}} & & #4 \\
        & \text{subject to~~}
        \ProcessList{#5}{ \insertopticonst }
        & & \nonumber
    \end{alignat}%
    }%
\endgroup%
}%
\newcommand\insertopticonst[1]{& & #1\\&}

\begin{document}

\title{Experimental entropic uncertainty relations in dimensions three to five}

\author{Laura Serino}
\affiliation{Paderborn University, Integrated Quantum Optics, Institute for Photonic Quantum Systems (PhoQS), Warburger Str.\ 100, 33098 Paderborn, Germany}

\author{Giovanni Chesi}
\affiliation{QUIT Group, Physics Department, Univ. Pavia, INFN Sez.~Pavia, Via Bassi 6, 27100 Pavia, Italy}

\author{Benjamin Brecht}
\affiliation{Paderborn University, Integrated Quantum Optics, Institute for Photonic Quantum Systems (PhoQS), Warburger Str.\ 100, 33098 Paderborn, Germany}

\author{Lorenzo Maccone}
\affiliation{QUIT Group, Physics Department, Univ. Pavia, INFN Sez.~Pavia, Via Bassi 6, 27100 Pavia, Italy}

\author{Chiara Macchiavello}
\affiliation{QUIT Group, Physics Department, Univ. Pavia, INFN Sez.~Pavia, Via Bassi 6, 27100 Pavia, Italy}

\author{Christine Silberhorn}
\affiliation{Paderborn University, Integrated Quantum Optics, Institute for Photonic Quantum Systems (PhoQS), Warburger Str.\ 100, 33098 Paderborn, Germany}

\begin{abstract}
We provide experimental validation of tight entropic uncertainty relations for the Shannon entropies of observables with mutually unbiased eigenstates in high dimensions. In particular, we address the cases of dimensions $d = 3$, $4$ and $5$ and consider from $2$ to $d+1$ mutually unbiased bases. The experiment is based on pulsed frequency bins measured with a multi-output quantum pulse gate, which can perform projective measurements on a complete high-dimensional basis in the time-frequency domain. Our results fit the theoretical predictions: the bound on the sum of the entropies is never violated and is saturated by the states that minimize the uncertainty relations.
\end{abstract}
\maketitle

\section{Introduction}
The uncertainty principle was originally formulated by Heisenberg \cite{heisenberg1927} to describe the fundamental limit on the precision of quantum measurements in terms of measurement disturbance, namely there exist measurements that irreversibly alter the state of the measured system.
It was later generalized by Robertson \cite{robertson1929} as a set of relations derived from the postulates of quantum mechanics and expressing measurement incompatibility, which describes sets of measurements that cannot be jointly performed. 
Notably, these two concepts remained distinct until very recently, when measurement incompatibility was shown to be a sufficient condition for disturbance \cite{erba2024}. 

The practical relevance of the uncertainty relations has historically evolved and grown with the development of modern quantum technologies. Today, in fact, establishing tight uncertainty bounds and identifying the states that saturate them are crucial for achieving the maximum precision limit in applications such as quantum cryptography, metrology and thermodynamics.
Towards this end, uncertainty relations for two observables have been reformulated in terms of the Shannon entropy \cite{hirshcman1957}, leading to the so-called entropic uncertainty relations (EURs) \cite{beckner1975, birula1975,deutsch1983,kraus1987,maassen1988,coles2017}. 
Unlike the former Robertson bound, the entropic bounds are state-independent. Moreover, the EURs can be expressed in terms of Rényi entropies \cite{maassen1988,coles2017}, which generalize Shannon's entropy, thus providing many different measures of uncertainty. 

The most well-known set of EURs is arguably the result by Maassen and Uffink \cite{maassen1988}, which quantifies the incompatibility between two observables.  
This bound is tight, meaning it can be saturated, if the two observables correspond to mutually unbiased bases (MUBs) \cite{wootters1988,durt2010}. 
Extending this bound beyond two MUBs is of both theoretical and practical interest, motivated by the role of high dimensions and multipartite systems in quantum information protocols, as well as by fundamental questions on the complementarity of quantum observables. 
Since every physical system admits at least three MUBs and, in prime-power dimensions $d$, $d+1$ MUBs are known to exist \cite{wootters1988,durt2010,brierley2010}, the total number of uncertainty relations at a specific dimension can grow very large, amounting to $2(2^d-1)-d$ in the latter case. 
Despite many efforts \cite{ivanovic1992,sanchez1995,azarchs2004,coles2017}, the analytic bounds found so far are either not tight or are tight only for specific dimensions and numbers of observables. However, numerical studies have obtained tight EUR bounds for up to $d+1$ MUBs in dimensions 3, 4 and 5 \cite{riccardi2017}. 

Experimentally, the traditional Robertson formulation of the uncertainty principle has been validated in a plethora of different contexts---from position-momentum \cite{shull1969,leavitt1969,nairz2002,guasti2022} and time-energy \cite{szriftgiser1996} relations to less conventional non-commuting observables such as number of Cooper pairs versus wavefunction phase in a superconductor \cite{elion1994}. These uncertainty relations are also implicitly tested every time the ultimate measurement precision bounds \cite{giovannetti2004,giovannetti2006} are assessed and applied \cite{lahaye2004,nagata2007,schliesser2009,cimini2018,tsarev2018,daryanoosh2018,saleem2024}. Furthermore, stronger uncertainty relations based on variances have been devised \cite{maccone2014,bagchi2016} and experimentally confirmed \cite{wang2016,ma2017,xiao2017,xiao2020}.

In contrast to the extensively verified Robertson formulation, entropic uncertainty relations have only more recently begun to be experimentally explored.
Experiments testing EURs have focused on the later formulation of uncertainty relations in terms of measurement noise and disturbance \cite{ozawa2003} and entropies conditioned on a quantum memory \cite{berta2010}. 
These were verified for neutronic spin systems \cite{erhart2012,sulyok2013,sulyok2015} and entangled photon pairs \cite{prevedel2011,li2011}. More recently, new uncertainty relations formulated in terms of relative entropy have been devised and experimentally validated \cite{barchielli2017,ding2020,liu2022}. However, to date, no experiment has directly tested the high-dimensional entropic bounds predicted in \cite{riccardi2017}. This gap largely reflects the experimental challenges of operating with discrete high-dimensional Hilbert spaces, which require simultaneous measurements in all elements of a high-dimensional basis \cite{yan2024}.

Here, we present the first experimental verification of all the EUR bounds predicted in \cite{riccardi2017} for dimensions 3, 4, and 5. We achieve this by encoding information in the time-frequency degree of freedom of photons, namely in pulsed frequency bins and their superpositions \cite{serino2025}. For this encoding, we can implement arbitrary projective measurements using a multi-output quantum pulse gate (mQPG) based on sum-frequency generation in a dispersion-engineered waveguide \cite{serino2023}. In particular, to verify the EUR bounds, we simultaneously project an input state onto all the elements of an arbitrary $d$-dimensional frequency-bin basis, selected through appropriate spectral shaping of the pump pulse driving the process.

Our experimental results verify the theoretically predicted entropic bounds for any number of MUBs in up to 5 dimensions, and confirm that these bounds are saturated by the states that, in Ref.~\cite{riccardi2017}, are found to minimize the EURs. 
These results complement our experimental investigation \cite{serino2024} of the effect of inequivalent classes of MUBs \cite{durt2010, brierley2010} on EURs in dimension 5.
Together, this work and \cite{serino2024} demonstrate the potential of our scalable setup to extend the verification of uncertainty relations to higher dimensions and detect further inequivalent sets of MUBs for $d>5$.

The remainder of this paper is structured as follows. In Section~\ref{2}, we summarize the main results from \cite{riccardi2017}. In Section~\ref{3} we describe how we implement a set of MUBs in high dimensions and the experiment verifying the numerical tight bounds on the EURs. Then, we discuss our results and compare them to the theoretical predictions from \cite{riccardi2017} and to previously known bounds. Finally, in Section~\ref{4}, we draw our conclusions.
    
\section{Framework}
\label{2}
Here, we concisely report the uncertainty relations for which we are going to provide experimental validations.
We focus on the formulation of EURs in terms of the Shannon entropy
\begin{equation}
    H(\hat{O}) = -\sum_{j}p_j\log_2p_j \,,
\end{equation}
where $p_j = |\langle o_j|\psi\rangle|^2$ are the Born probabilities for the measurement of the observable $\hat{O}$, with eigenvectors $\{|o_j\rangle\}_j$, on a state $|\psi\rangle$. 

In terms of the Shannon entropies of two observables $\hat{A}$ and $\hat{B}$, the Maassen-Uffink \cite{maassen1988} bound reads
\begin{equation} \label{mu}
    H(\hat{A}) + H(\hat{B}) \geq -\log{c}\,,
\end{equation}
where $c \equiv \text{max}_{j,k}|\langle a_k| b_j\rangle|^2$ is the maximum overlap between the eigenstates of $\hat{A}$ and $\hat{B}$, respectively $\{\ket{a_k}\}_k$ and $\{\ket{b_j}\}_j$.
In a discrete Hilbert space of dimension $d$, this bound is tight, meaning that it can be saturated, if the eigenstates of $\hat{A}$ and $\hat{B}$ are mutually unbiased bases (MUBs) \cite{wootters1988,durt2010}, i.e., if $|\langle a_k| b_j\rangle| = 1/\sqrt{d} \quad \forall \, j,k\in [0,d)$.

Given a set of $m$ observables $\{\hat{M}_k\}_{k=l_1}^{l_m}$ in dimension $d$ with mutually unbiased eigenstates, the general expression of the addressed inequalities reads
\begin{equation} \label{geneur}
    \sum_{k=l_1}^{l_m}H(\hat{M}_k) \geq \mathcal{B}_{d,m}\,,
\end{equation}
namely, the lower bound $\mathcal{B}$ on the sum of the Shannon entropies of the observables is specific for each $d$ and $m$ considered. However, in general, it is not unique: as mentioned in the Introduction, in the particular case with $d = 5$ and $m = 3$, there are two distinct bounds depending on the choice of the triple $\{l_1, l_2, l_3\}$. In our notation, $l_j$ identifies the $j$-th MUB in the set $\{l_j\}_{j=1}^m$.

We express the MUBs in terms of Hadamard matrices, following the construction in Refs.~\cite{durt2010,brierley2010}. The explicit form is reported in Appendix~\ref{app}. There, every matrix represents a basis, each column being a state of the pertaining basis, and $l_j \in \{A,B,...\}\,\,\, \forall\, j\in [1,m]$.
We report in Table~\ref{bounds} the lower bounds of the uncertainty relations in Eq.~\eqref{geneur}.

\begin{table}
\centering
\begin{tabular}{|c|c|c|c|c|c|} 
\cline{2-6}
\multicolumn{1}{c|}{} & $m=2$ & $m=3$ & $m=4$ & $\,m=5\,$ & $m=6$ \\
\cline{2-6}
\hline
$d=3$ & $\log_2 3$ & 3 & 4 & - & - \\
$d=4$ & 2 & 3 & 5 & 7 & - \\
$d=5$ & $\log_2 5$ & $4.43 \vee 2\log_2 5$ & 6.34 & 8.33 & 10.25 \\
\hline
\end{tabular}
\caption{Lower bounds $\mathcal{B}$ of the EURs in Eq.~\ref{geneur} for dimension $d$ three to five and $m$ MUBs.}
\label{bounds}
\end{table}

Firstly, we note that, if $m = 2$, then the Maassen-Uffink relation in Eq.~\eqref{mu} holds in any dimension, implying that the uncertainty is minimized by the eigenstates of the addressed observables, i.e., the states of the pertaining MUBs. Henceforth, we will inspect the saturation of the uncertainty relations in Eq.~\eqref{geneur} for each dimension $d\in[3,5]$ with $m>2$.

In the case $d=3$, the set of states $|\psi\rangle_{\rm opt}$ that saturates Eq.~\eqref{geneur} is the same for both $m=3$ and $m=4$ and is given by
\begin{equation} \label{optd3}
    |\psi\rangle_{\rm opt} = \frac{|j_1\rangle + e^{i\phi}|j_2\rangle}{\sqrt{2}}
\end{equation}
with $j_1, j_2 = 0,1,2$, $j_1 \neq j_2$ and $\phi = \pi/3, \pi, 5\pi/3$.

In dimension $d = 4$, the triples of Hadamard matrices that are mutually unbiased belong to a three-parameter family \cite{brierley2010,durt2010}. However, there is a unique value for each of these parameters such that four and five MUBs can be constructed. We considered this case, where the complete set of MUBs can be obtained, which is reported in Appendix~\ref{d4}. Differently from the case $d=3$, the states saturating Eq.~(\ref{geneur}) depend on the choice of the observables \cite{riccardi2017}. However, the states that minimize the uncertainty of three and four observables, also minimize the one of the complete set of MUBs. For $m = 3$, if one of the MUBs is the computational basis ($A$ in Appendix~\ref{d4}), then the expression of the optimal states is the same as the one in Eq.~(\ref{optd3}), where the parameters $j_1, j_2 \in [0,3]$ and $\phi \in \{\pm k\pi/2\}_{k=0}^2$ depend on the choice of the other two bases. We report in Table~\ref{d4optst1} the values of the parameters for this case. Conversely, if $A \notin \{l_1,l_2,l_3\}$, the optimal states, in terms of the eigenstates of the computational basis, read
\begin{equation} \label{optd32}
    |\psi\rangle_{\rm opt} = \frac{1}{2}\sum_{j=0}^3e^{i\phi_j}|j\rangle,
\end{equation}
namely, they are defined by three of the four phases $\phi_j$. In Table~\ref{d4optst2} we fix the phase $\phi_0 = 0$ and show the relative phases of the optimal states for each triple $\{l_1,l_2,l_3\}$. In the case $m=4$, the states that are optimal for a set $\{\bar{l}_1,\bar{l}_2,\bar{l}_3,\bar{l}_4\}$ are the same that optimize the corresponding triplet $\{\bar{l}_1,\bar{l}_2,\bar{l}_3\}$, being $\bar{l}_4$ one of the two bases left. It is worth noting that the optimal states in Eq.~(\ref{optd32}) can be recast as the expression in Eq.~(\ref{optd3}), namely as the superposition of two states, by expanding $|\psi\rangle_{\rm opt}$ over a different basis. Therefore, the set of optimal states identified in Eq.~(\ref{optd32}) arises from our convention of adopting the computational basis to define $|\psi\rangle_{\rm opt}$, and the fundamental structure is the one in Eq.~(\ref{optd3}), with $|j_1\rangle$ and $|j_2\rangle$ states belonging to any of the five MUBs. 

\begin{center}
\begin{table}
\begin{tabular}{ |c|c|c|c|c| } 
\hline
$\{l_1,l_2,l_3\}$ & $\{l_1,l_2,l_3,l_4\}$ & $j_1$ & $j_2$ & $\phi$ \\
\hline
\hline
\multirow{2}{2.5em}{$ABC$} & \multirow{2}{7.5em}{$ABCD$, $ABCE$} & 0 & 1 & $0, \pi$ \\ 
 & & 2 & 3 & $0, \pi$ \\  
\hline
\multirow{2}{2.5em}{$ABD$} & \multirow{2}{7.5em}{$ABCD$, $ABDE$} & 0 & 2 & $0, \pi$ \\ 
& & 1 & 3 & $0, \pi$ \\
\hline
\multirow{2}{2.5em}{$ABE$} & \multirow{2}{7.5em}{$ABCE$, $ABDE$} & 0 & 3 & $0, \pi$ \\ 
& & 1 & 2 & $0, \pi$ \\
\hline
\multirow{2}{2.5em}{$ACD$} & \multirow{2}{7.5em}{$ABCD$, $ACDE$} & 0 & 3 & $\pm \pi/2$ \\ 
& & 1 & 2 & $\pm \pi/2$ \\
\hline
\multirow{2}{2.5em}{$ACE$} & \multirow{2}{7.5em}{$ABCE$, $ACDE$} & 0 & 2 & $\pm \pi/2$ \\ 
& & 1 & 3 & $\pm \pi/2$ \\
\hline
\multirow{2}{2.5em}{$ADE$} & \multirow{2}{7.5em}{$ABDE$, $ACDE$} & 0 & 1 & $\pm \pi/2$ \\ 
& & 2 & 3 & $\pm \pi/2$ \\
\hline
\end{tabular}
\caption{Optimal states $|\psi\rangle_{\rm opt}$ in Eq.~(\ref{optd3}) saturating the bounds $\mathcal{B}_{4,3}$ and $\mathcal{B}_{4,4}$ in the EURs of Eq.~(\ref{geneur}) for triples of MUBs including the computational basis $A$ (first column) and for the corresponding quadruples (second column).}
\label{d4optst1}
\end{table} 
\end{center}

\begin{center}
\begin{table}
\begin{tabular}{ |c|c|c|c|c| } 
\hline
$\{l_1,l_2,l_3\}$ & $\{l_1,l_2,l_3,l_4\}$ & $\phi_1$ & $\phi_2$ & $\phi_3$ \\
\hline
\hline
\multirow{4}{2.5em}{$BCD$} & \multirow{4}{7.5em}{$ABCD$, $BCDE$} & $\pi/2$ & $\pi/2$ & $0$ \\ 
& & $-\pi/2$ & $-\pi/2$ & $0$ \\  
& & $\pi/2$ & $-\pi/2$ & $\pi$ \\ 
& & $-\pi/2$ & $\pi/2$ & $\pi$ \\  
\hline
\multirow{4}{2.5em}{$BCE$} & \multirow{4}{7.5em}{$ABCE$, $BCDE$} & $\pi/2$ & 0 & $\pi/2$ \\ 
& & $-\pi/2$ & 0 & $-\pi/2$ \\
& & $\pi/2$ & $\pi$ & $-\pi/2$ \\ 
& & $-\pi/2$ & $\pi$ & $\pi/2$ \\
\hline
\multirow{4}{2.5em}{$BDE$} & \multirow{4}{7.5em}{$ABDE$, $BCDE$} & $0$ & $\pi/2$ & $\pi/2$ \\ 
& & $0$ & $-\pi/2$ & $-\pi/2$ \\
& & $\pi$ & $\pi/2$ & $-\pi/2$ \\ 
& & $\pi$ & $-\pi/2$ & $\pi/2$ \\
\hline
\multirow{4}{2.5em}{$CDE$} & \multirow{4}{7.5em}{$ACDE$, $BCDE$} & $\pi$ & 0 & 0 \\ 
& & 0 & $\pi$ & 0 \\
& & 0 & 0 & $\pi$ \\
& & $\pi$ & $\pi$ & $\pi$ \\
\hline
\end{tabular}
\caption{Optimal states $|\psi\rangle_{\rm opt}$ in Eq.~(\ref{optd32}) saturating the bounds $\mathcal{B}_{4,3}$ and $\mathcal{B}_{4,4}$ in the EURs of Eq.~(\ref{geneur}) for triples of MUBs not including the computational basis $A$ (first column) and for the corresponding quadruples (second column). We set $\phi_0 = 0$.}
\label{d4optst2}
\end{table} 
\end{center}

In dimension $d = 5$, there are two inequivalent sets of triples of MUBs \cite{brierley2010,durt2010}, implying the two distinct bounds in Table~\ref{bounds} for the case $m = 3$ \cite{serino2024}. The states achieving the bound $\mathcal{B}_{5,3}^{(1)} = 2\log_25\simeq 4.64$ are the eigenstates of the observables involved in the uncertainty relation. The lower bound $\mathcal{B}_{5,3}^{(2)} \simeq 4.43$ featured by one of the two inequivalent sets, is obtained with the states
\begin{equation} \label{optd5}
    |\psi\rangle_{\rm opt} = \sum_{j=0}^{4}\psi_je^{i\phi_j}|j\rangle
\end{equation}
such that one of the coefficients $\psi_j$ is null and the four coefficients left are pairwise equal. In Section~\ref{3}, we provide experimental proof only of the bound $\mathcal{B}^{(1)}_{5,3}$ corresponding to the MUB set analyzed in \cite{riccardi2017}. The second, inequivalent bound $\mathcal{B}^{(2)}_{5,3}$ was recently identified and verified in \cite{serino2024}, where we also inspect the pertaining minimum uncertainty states.

For $m > 3$, there is a single equivalence class of sets of MUBs \cite{brierley2010}. Thus, for each choice of the observables $\{\hat{M}_k\}_{k=l_1}^{l_m}$ in Eq.~\ref{geneur} we have a single bound $\mathcal{B}_{5,m}$. However, we have, again, that the uncertainty of different sets of observables is minimized by different states. With $m = 4$ and $m = 5$, the structure of $|\psi\rangle_{\rm opt}$ is the same as the one in Eq.~\ref{optd5}. We found and experimentally tested nine optimal states in the case $m = 4$ for the MUBs $ABCD$ and five in the case $m = 5$ for $ABCDE$. In the former scenario, the states $|\psi\rangle_{\rm opt}$ are defined by coefficients $\psi_j \in \{0,0.19,0.68\}$ and phases $\phi_j \in \{\pm k\pi/5\}_{k=1}^5$, while, in the latter, by $\psi_j \in \{0,0.11,0.70\}$ and the same set of phases.
Finally, in the case $m = 6$, the optimal states are given again by Eq.~\ref{optd3}, with $j_1, j_2 \in [0,4]$ and $\phi_j \in \{\pm (2k+1)\pi/5 \}_{k=0}^2$.

\section{Experimental verification of the EURs with pulsed frequency bins}
\label{3}
\subsection{Experiment}
In the proposed experiment, we consider a Hilbert space constructed from pulsed frequency bins—-an encoding alphabet based on the time-frequency degree of freedom of photons \cite{brecht2015}. 
The $d$-dimensional computational basis $A$ consists of $d$ broadband Gaussian frequency bins, each centered at a distinct frequency. 
The bases mutually unbiased to $A$ are then generated by superimposing these frequency bins with phase relationships dictated by the Hadamard matrices $\{B,C,...\}$  in Appendix~\ref{app} \cite{lu23b, widomski23, serino2025}.

Experimentally, we can estimate the probability distribution required to calculate the entropy by normalizing the counts obtained from projecting multiple copies of the input states onto the selected MUBs.
Thus, we require a device capable of simultaneously projecting states from this $d$-dimensional Hilbert space onto all elements of an arbitrarily chosen basis.
In the time-frequency domain, this can be achieved using a multi-output quantum pulse gate (mQPG) \cite{serino2023, serino2025}, a high-dimensional decoder for time-frequency pulsed modes based on sum-frequency generation in a dispersion-engineered waveguide. 
The projection basis is determined by shaping the spectrum of the pump pulse driving the process, and the outcome of the projection onto each basis element corresponds to a probability of detecting a click in a specific output channel, each defined by a distinct output frequency.

The mQPG operation is described by a positive-operator-valued measure (POVM) $\{\pi^\gamma\}$, where each POVM element $\pi^\gamma$ describes the measurement operator of a single channel set to project onto mode $\gamma$ \cite{brecht2014, ansari2017, serino2023}. Under ideal conditions, each channel performs perfect single-mode projections $\pi^\gamma=\ketbra{\gamma}$. However, experimental imperfections can lead to systematic errors in the POVMs, which then are described by the more general expression $\pi^\gamma=\sum_{ij}m^\gamma_{ij}\ket{a_i}\bra{a_j}$, with $\ket{a_i}$ and $\ket{a_j}$ eigenstates of the computational basis. When measuring a pure $d$-dimensional input state $\rho^\xi=\ketbra{\xi}$, we will obtain output $\gamma$ with probability $p^{\gamma\xi} = \mathrm{Tr}(\rho^\xi \pi^\gamma)$.

\begin{figure}
	\centering
	\includegraphics{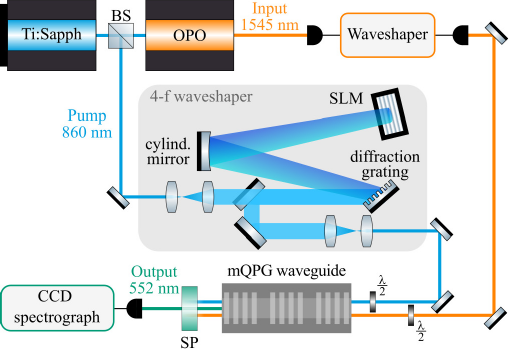}
	\caption{Schematic of the experimental setup. The signal and pump pulses, centered at \SI{1545}{\nm} and \SI{860}{\nm} respectively, are generated by a combination of a Ti:Sapphire ultrafast laser with an optical parametric oscillator (OPO). The signal pulse is shaped by a commercial waveshaper (Finisar 4000S) to generate the frequency-bin states in input. The remaining pump pulse, separated by a beam splitter (BS), is shaped as the full measurement basis using an in-house-built 4f-waveshaper consisting of a diffraction grating, a cylindrical mirror and a spatial light modulator (SLM). Both beams are coupled into the mQPG waveguide, where the signal is up-converted into the different output frequencies as a result of the mQPG projections. The output beam, centered at around \SI{552}{nm}, is isolated using a short-pass (SP) filter and then detected by a commercial CCD spectrograph (Andor Shamrock 500i).}
	\label{fig:setup}
\end{figure}

A schematic of the experimental setup is shown in Figure \ref{fig:setup}. The signal and pump pulses, centered at \SI{1545}{\nm} and \SI{860}{\nm} respectively, are generated by a combination of a Ti:Sapphire ultrafast laser with an optical parametric oscillator (OPO) at a repetition rate of \SI{80}{\MHz}. The signal pulse is shaped by a commercial waveshaper (Finisar 4000S) to generate the frequency-bin states in input, whereas the pump pulse is shaped by an in-house-built 4f-waveshaper to generate the $d$-dimensional frequency-bin basis for the measurement, with $d\in\{3,4,5\}$. Both beams are coupled into the mQPG waveguide, where the signal modes are up-converted into a different output channel based on their overlap with each pump mode, i.e., with each eigenstate of the chosen measurement basis. The output beam, centered at around \SI{552}{nm}, is separated from the unconverted signal and pump beams by a shortpass filter and then detected by a commercial CCD spectrograph (Andor Shamrock 500i). This frequency-resolved detection allows us to separate the output channels, each centered at a distinct frequency. The number of counts detected at each output frequency indicates the number of photons measured in the corresponding mode.

For the experimental verification of the EURs found in \cite{riccardi2017} and in this work, we measure the entropy of a large sample of states for different combinations of $m$ observables $\{\hat{M}_k\}_{k=l_1}^{l_m}$, with $2\leq m \leq d+1$. We group the probed input states into four categories: \emph{optimal states}, found in Ref.~\cite{riccardi2017}, which saturate the aforementioned entropy bounds; \emph{internal} and \emph{external eigenstates}, which are, respectively, the eigenstates of an observable in the set $\{\hat{M}_k\}_{k=l_1}^{l_m}$ appearing in the addressed uncertainty relation or of one of the complementary observables; \emph{random states} generated by randomly sampling amplitude and phase coefficients from a uniform distribution. Although this method does not allow for truly uniform sampling of the parameter space, this is not relevant in the scope of this work, which concerns only the lower bound of the joint entropy distribution.

\subsection{Results and discussion}
\begin{figure}
    \centering
    \includegraphics{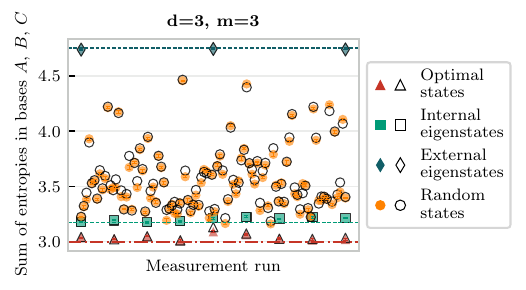}
    \caption{Sum of the entropy calculated in the MUBs $A$, $B$, $C$ in $d$=3 for different types of input states: optimal states (red triangles), states from bases $A$, $B$, or $C$ (green squares), states from the last MUB $D$ (blue diamonds) and random states (yellow circles). The filled markers show the experimental data, whereas the hollow markers describe the predicted results based on the characterized imperfect POVMs. The dashed blue and green lines show the predicted entropy for the external and internal eigenstates, respectively. The red dash-dotted line indicates the EUR bound found in \cite{riccardi2017} for $d=3$ and $m=3$, which is saturated by the optimal states.}
    \label{fig:scatterplot}
\end{figure}

\begin{figure*}
    \centering
    \includegraphics{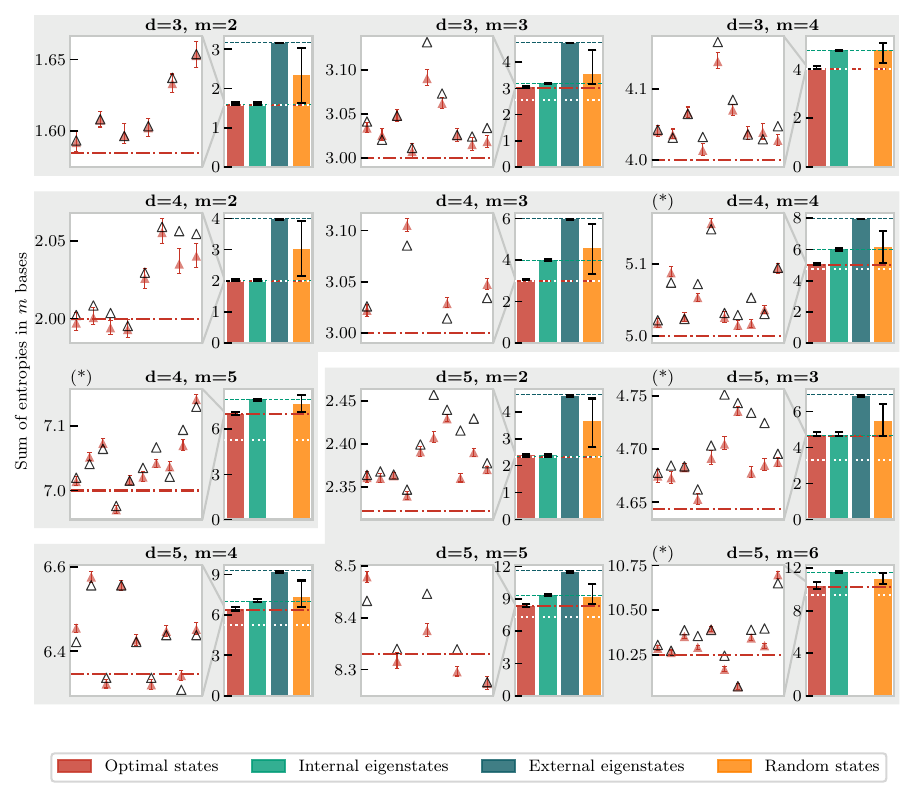}
    \caption{Measured sum of entropy in $m$ MUBs for different dimensions $d$. For each combination of $m$ and $d$, we only show the results for a single set of bases. Left: scatterplot showing the entropy of individual optimal states (filled triangles) and corresponding simulations (hollow triangles) in an enlarged scale. In the plots marked by (*) we only show the first ten optimal states out of a larger number of data. Right: barplot showing the mean value and the total spread of the sum of entropy for different sets of input states: from left to right, optimal states (red), internal eigenstates (green), external eigenstates (blue) and random states (yellow). The dashed blue and green lines show the predicted entropy for the external and internal eigenstates, respectively. The red dash-dotted line indicates the EUR bounds found in \cite{riccardi2017} and in \cite{serino2024}, which is saturated by the optimal states. In contrast, the white dotted lines show the previously known EUR bounds prior to \cite{riccardi2017}. See the text for more information.
    }
    \label{fig:results}
\end{figure*}

For each probed input state and measurement basis, we calculate the entropy by estimating the probability $p_i$ of each measurement outcome as the normalized counts in the corresponding channel. We obtain distributions of entropy values such as the one in Figure \ref{fig:scatterplot}, which shows the sum of the entropy calculated in the first three MUBs in Appendix~\ref{app} for different input states from a three-dimensional Hilbert space. For each data point, the error bars indicate the 10\%-90\% spread of the entropy distribution obtained from a Monte Carlo simulation based on 500 samples extracted from the measured count statistics. The error bars are not visible in most points due to their narrow extent, indicating a small statistical error. 

Figure \ref{fig:results} summarizes the experimental results for the sum of entropies $\sum_{k=l_1}^{l_m}H(\hat{M}_k)$ of $d$-dimensional input states measured over a set $\{\hat{M}_k\}_{k=l_1}^{l_m}$ of $m$ MUBs, for all combinations of $d\in\{3,4,5\}$ and $m\in\{3,\dots, d+1\}$. The individual measurements of the optimal states are highlighted in the scatterplots in an enlarged scale. We only show the measurement results for a single set of bases for each combination of $d$ and $m$, as the observed lower bounds are homogeneous over all sets of MUBs with the only exception of the case of $d=5$ and $m=3$.
For this particular combination, in fact, we only considered a set of MUBs equivalent to the one in Ref.~\cite{riccardi2017}, and we verified that the EUR bound is saturated by the eigenstates of those MUBs, as expected. The case of the inequivalent set of MUBs, which leads to a different EUR bound, is analyzed in depth and experimentally verified in Ref.~\cite{serino2024}.

For each combination of $d$ and $m$, the barplot shows the mean value of the sum of entropies for different sets of input states: from left to right, optimal states (red), internal eigenstates 
(green), external eigenstates 
(blue) and random states (yellow). The error bars mark the total spread of the data distribution. 
The results show that the optimal states saturate the entropy bounds (red dashdotted lines), while the internal eigenstates---with some exceptions---and external eigenstates do not, sitting at $(m-1) \log_2{d}$ (green dashed line) and $m \log_2{d}$ (blue dashed line), respectively. The exceptions are the cases with $m=2$ and $(d=5\land m=3)$, in which the internal eigenstates coincide with the optimal states, and the EUR bound is exactly $(m-1) \log_2{d}$.
In the case of random states, the mean sum of the entropies is always above the lower bound, as expected.

In the barplot, we also show (as dotted white lines) the analytic lower bounds known prior to the tight numeric ones of Ref.~\cite{riccardi2017}. For $m = 2$, the bound from the Maassen-Uffink uncertainty relation \cite{maassen1988} is always tight. For $2 < m < d+1$, the bound found in Ref.~\cite{azarchs2004} is tight only in the case $(d = 4 \land m = 3)$. While this is still the best analytic bound for $(d=3 \land m=3)$, $(d=4 \land m=4)$, $(d=5\land m=3)$, $(d=5 \land m=4)$ and $(d=5 \land m=5)$, it fails to saturate the EURs. The bound found in Ref.~\cite{ivanovic1992} for $m=d+1$ is tight in the case $(d=3 \land m=4)$, but it is weak for $(d=4 \land m=5)$ and $(d = 5 \land m=6)$. We find that our experimental data confirms these predictions, reaching the bound from \cite{riccardi2017} but not the previous bound when it is not tight.

In general, all data sets match very closely the theoretical predictions, with only minor discrepancies. However, one could be concerned that in some cases the statistical error bar is significantly smaller than this discrepancy, resulting in some data points that appear to violate the lower entropy bound.
The small difference between the theoretical and measured entropy values can be explained considering the systematic error in the detection system. To verify this, we characterized the performance of the mQPG-based decoder through a quantum detector tomography \cite{lundeen2009, ansari2018, serino2023} to reconstruct the true POVMs. We observed a measurement error (cross-talk) ranging from 0.1\% to 2\%, depending on the considered basis and dimension, with larger $d$ generally leading to higher errors. From the reconstructed POVMs, we calculated the expected entropy that one would observe measuring the probed input states with this imperfect system, obtaining the hollow markers in Figure \ref{fig:scatterplot}. These estimates very closely match the measured values, confirming that discrepancies with the theoretical predictions are to be attributed to the detection system.

\section{Conclusions}
\label{4}
We provided the first experimental verification of the EURs for dimensions 3, 4 and 5, introduced in Ref.~\cite{riccardi2017}, in the Hilbert space formed by pulsed frequency bins and their superpositions. By performing high-dimensional time-frequency projections through an mQPG, we have probed several input states, measured their superpositions with the MUBs and retrieved the sum of the Shannon entropies of the pertaining observables $\{\hat{M}_k\}_{k=l_1}^{l_m}$ for every number of MUBs $m\in[2, d+1]$. In the case $m=2$, our results reproduce the Maassen-Uffink uncertainty relations for Shannon entropies. Where $m > 2$, we verified that the optimal states described in Ref.~\cite{riccardi2017} minimize the EURs for each $d$ and $m$, and that the bounds assessed there are not achived nor violated by any other input state.
Simulations based on the reconstructed POVMs of the detector have shown that the small discrepancies from the theory are explained by systematic measurement errors.

The key element that enabled this experimental verification resided in the capability of the mQPG to perform simultaneous projections onto all elements of the selected basis. It is worth noting that, while this work focused on the EUR bounds in dimension up to $d=5$ shown in Ref.~\cite{riccardi2017}, the experimental setup can be straightforwardly adapted to operate in larger dimensions \cite{serino2025, de2024}, facilitating the experimental verification of possible EUR bounds in higher-dimensional Hilbert spaces and even enabling the detection of inequivalent sets of MUBs \cite{serino2024}.

These results mark an important step toward the realization of quantum communication systems in high dimensions. A remarkable example is the one of discrete-variable quantum key distribution, where high-dimensional encodings offer increased secret key rates and resilience to noise \cite{sheridan2010}, and unconditional security in real-world implementations can be guaranteed if the number of secret bits of the key is bounded via suitable EURs \cite{tomamichel2011}. 
Additionally, extending the EURs tested in this work to entropies conditioned on a quantum memory \cite{berta2010} will enable novel entanglement-based high-dimensional QKD protocols, which can be implemented using a scheme similar to the one employed here \cite{brecht2015}.

\section{Acknowledgements}
The authors acknowledge support from the EU H2020 QuantERA ERA-NET Cofund in Quantum Technologies project QuICHE. 
G.C.\ acknowledges support from from the PNRR MUR Project PE0000023-NQSTI. 
C.M.\ acknowledges support from the PRIN MUR Project
2022SW3RPY. 
L.M.\ acknowledges support from the PRIN MUR Project 2022RATBS4 and from the U.S.\ Department of Energy, Office of Science, National Quantum Information Science Research Centers, Superconducting Quantum Materials and Systems Center (SQMS) under Contract No.\ DE-AC02-07CH11359.

\clearpage\appendix

\section{Explicit expression of the MUBs in terms of Hadamard matrices} \label{app}
Following Ref.~\cite{brierley2010}, here we report the Hadamard matrices defining MUBs in dimensions 3, 4 and 5. Each matrix represents a basis, where the columns are the pertaining orthonormal states.
\setlength{\abovedisplayskip}{0pt}
\subsection{$d=3$} \label{d3}
Complete set of MUBs in dimension $d=3$.
\\
\begin{equation}
\begin{aligned}
&A = \begin{pmatrix}
1 & 0 & 0 \\
0 & 1 & 0 \\
0 & 0 & 1
\end{pmatrix} \quad  &B = \frac{1}{\sqrt{3}} \begin{pmatrix}
1 & 1 & 1 \\
1 & \omega & \omega^2 \\
1 & \omega^2 & \omega
\end{pmatrix} \\ 
&C =  \frac{1}{\sqrt{3}}\begin{pmatrix}
1 & 1 & 1 \\
\omega^2 & \omega & 1 \\
1 & \omega & \omega^2
\end{pmatrix} \quad &D = 
\frac{1}{\sqrt{3}}\begin{pmatrix}
1 & 1 & 1 \\
\omega & \omega^2 & 1 \\
1 & \omega^2 & \omega \nonumber
\end{pmatrix}
\end{aligned}
\end{equation}
with
\begin{equation}
 \omega\equiv \exp\left(i\frac{2}{3}\pi\right). \nonumber 
\end{equation}

\subsection{$d=4$} \label{d4}
Complete set of MUBs in dimension $d=4$.
\\
\begin{equation*}
\begin{aligned}
&A = \begin{pmatrix}
1 & 0 & 0 & 0 \\
0 & 1 & 0 & 0 \\
0 & 0 & 1 & 0 \\
0 & 0 & 0 & 1
\end{pmatrix} \,\,\,
B =  \frac{1}{2} \begin{pmatrix}
1 & 1  & 1   & 1  \\
1 & 1  & -1  & -1 \\
1 & -1 & -1  & 1   \\
1 & -1 & 1   & -1
\end{pmatrix} \\ 
&C=  \frac{1}{2} \begin{pmatrix}
1  & 1  & 1  & 1        \\
1  & 1  & -1 & -1  \\
-i & i  & i  & -i        \\
i  & -i & i  & -i 
\end{pmatrix} \\
&D = \frac{1}{2} \begin{pmatrix}
1  & 1  & 1  & 1        \\
i  & -i & i  & -i   \\
-1 & -1 & 1  & 1  \\
i  & -i & -i & i  \\
\end{pmatrix} \\
&E = \frac{1}{2} \begin{pmatrix}
1  & 1  & 1  & 1    \\
i  & -i & i  & -i   \\
i  & -i & -i & i \\
-1 & -1 & 1  & 1
\end{pmatrix}
\end{aligned}
\end{equation*}

\subsection{$d=5$} \label{d5}
Complete set of MUBs in dimension $d=5$.
\\
\begin{equation*} \label{mub5}
\begin{aligned}
&A = \begin{pmatrix}
1 & 0 & 0 & 0 & 0 \\
0 & 1 & 0 & 0 & 0 \\
0 & 0 & 1 & 0 & 0 \\
0 & 0 & 0 & 1 & 0 \\
0 & 0 & 0 & 0 & 1
\end{pmatrix} \quad B = \frac{1}{\sqrt{5}} \begin{pmatrix}
1 & 1 & 1 & 1 & 1 \\
1 & \omega & \omega^2 & \omega^3 & \omega^4 \\
1 & \omega^2 & \omega^4 & \omega & \omega^3 \\
1 & \omega^3 & \omega & \omega^4 & \omega^2 \\
1 & \omega^4 & \omega^3 & \omega^2 & \omega 
\end{pmatrix} \\ 
&C = \frac{1}{\sqrt{5}} \begin{pmatrix}
1        & 1        & 1        & 1        & 1 \\
\omega   & \omega^2 & \omega^3 & \omega^4 & 1 \\
\omega^4 & \omega   & \omega^3 & 1        & \omega^2 \\
\omega^4 & \omega^2 & 1        & \omega^3 & \omega   \\
\omega   & 1        & \omega^4 & \omega^3 & \omega^2 
\end{pmatrix} \\
&D = \frac{1}{\sqrt{5}} \begin{pmatrix}
1        & 1        & 1        & 1        & 1 \\
\omega^3 & \omega^4 & 1        & \omega   & \omega^2 \\
\omega^2 & \omega^4 & \omega   & \omega^3 & 1       \\
\omega^2 & 1        & \omega^3 & \omega   & \omega^4   \\
\omega^3 & \omega^2 & \omega   & 1        & \omega^4 
\end{pmatrix} \\
&E = \frac{1}{\sqrt{5}} \begin{pmatrix}
1        & 1        & 1        & 1        & 1 \\
\omega^2 & \omega^3 & \omega^4 & 1        & \omega \\
\omega^3 & 1        & \omega^2 & \omega^4 & \omega \\
\omega^3 & \omega   & \omega^4 & \omega^2 & 1   \\
\omega^2 & \omega   & 1        & \omega^4 & \omega^3 
\end{pmatrix} \\
&F = \frac{1}{\sqrt{5}} \begin{pmatrix}
1        & 1        & 1        & 1        & 1 \\
\omega^4 & 1        & \omega   & \omega^2 & \omega^3 \\
\omega   & \omega^3 & 1        & \omega^2 & \omega^4 \\
\omega   & \omega^4 & \omega^2 & 1        & \omega^3   \\
\omega^4 & \omega^3 & \omega^2 & \omega   & 1 
\end{pmatrix}
\end{aligned}
\end{equation*}
with
\begin{equation}
    \omega\equiv \exp\left(i\frac{2}{5}\pi\right). \nonumber
\end{equation}
\newpage
\bibliographystyle{apsrev4-1}
\bibliography{bib}

\end{document}